\documentclass[aps,prd,twocolumn,nofootinbib]{revtex4-2}

\usepackage[colorlinks]{hyperref}
\usepackage{amssymb,amsmath}
\usepackage{graphicx}
\usepackage{subcaption}

\begin{document}
\title{Captured molecules could  make a Bose star visible}
\author{V. V. Flambaum}\email{v.flambaum@unsw.edu.au}
\author{I. B. Samsonov}\email{igor.samsonov@unsw.edu.au}
\affiliation{School of Physics, University of New South Wales, Sydney 2052, Australia}

\begin{abstract}
A Bose star passing through cold molecular clouds may capture atoms, molecules and dust particles. The observational signature of such an event would be a relatively small amount of matter that is gravitationally bound. This binding may actually be provided by invisible dark matter forming the Bose star. 
We may expect a relative excess of heavier atoms, molecules, and solid dust compared to the content of giant cold molecular clouds since the velocity of heavy particles at a given temperature is lower and it may be small compared to the escape velocity, $v_\mathrm{rms} = \sqrt{3k_\mathrm{B} T/m_\mathrm{gas}} \ll v_\mathrm{esc}=\sqrt{2GM/R}$. Finally, the velocity of this captured matter cloud may correlate with the expected velocity of free dark matter particles (e.g.\ expected  axion wind velocity relative to Earth).
\end{abstract}

\maketitle

\section{Introduction}
Bose stars  are assumed to be formed from cold dark matter (DM) fields (scalars, axions, etc.) which may form Bose condensate and solitons bound mainly by the gravitational  attraction \cite{Ruffini1969,Tkachev1986}. Bose condensation temperature is $ T_B \propto n^{2/3}/m$, so in thermal equilibrium axions (or scalars) form Bose condensate for sufficiently small mass $m$  and large  density $n$. Bose stars may comprise a significant fraction of dark matter which may be potentially detectable, as will be argued in this work.

A simple estimate, similar to that for hydrogen atom,  makes relation between mass $M\approx N m$ and size $R$  of Bose star, where $N$ is the number of bosons and $m$ is their mass. 
Estimating kinetic energy of the ground state using uncertainty relation for minimal momentum $p  \sim \hbar/r$ and estimating gravitational energy we obtain equation for the non-relativistic energy of Bose star: 
\begin{equation}
E=a_1 N\frac{\hbar^2}{mr^2} - a_2 \frac{N(N-1)}{2} \frac{Gm^2}{r},
\end{equation}
 where $a_1$, $a_2$ are constants $\sim 1$. Minimizing energy by varying  size $r$ or using virial theorem ($U=- 2E_k$, $E=-E_k$) we obtain size of gravitationally bound  Bose star   
\begin{equation}
R=\frac{4 a_1  \hbar^2}{a_2 (N-1)G m^3},
\end{equation}
 binding energy
\begin{equation}
E=N (N-1)^2  \frac{a_2^2 G^2 m^5}{16 a_1\hbar^2},
\end{equation}
and number density 
\begin{equation}
n  \sim \frac{N}{V}  \sim 0.004 N (N-1)^3  \frac{G^3 m^9}{\hbar^6},
\end{equation}
where the star volume is $V \sim 4 \pi R ^3/3$.
We see very strong dependence on the boson mass $m$. We also see that the density $n$ in   the ground state is proportional to $N^4$ so any, even small non-gravitational self interaction between bosons may become important  for large $N$.
We may add the self-interaction  contribution of the self-interaction term $\lambda \phi^4$ to the Bose star energy 
\begin{equation}
E=a_1 N\frac{\hbar^2}{mr^2} - a_2 \frac{N(N-1)}{2} \frac{Gm^2}{r}+a_3  \lambda \frac{N(N-1)}{2 m^2 r^3}  ,
\end{equation}
where $a_1$, $a_2$ $a_3$ are constants $\sim 1$. Using $dE/dr=0$,  this expression  gives quadratic equation for $r$ with two solutions. For the QCD axion, the selfinteraction constant $\lambda= -m^2/(24 f_a^2)$ is expressed via the axion decay constant $f_a$ (see e.g. \cite{Levkov2018,MarshReview}).  In the case of $\lambda<0$ the energy tends to minus infinity for $r \to 0$, so the Bose star is metastable. Solution with larger $r$ corresponds to a minimum whereas the smaller $r$ solution is maximum. More accurate analytical calculations using variational approach will be given in the next section.

Distribution of gas particles of ordinary matter is given by Boltzmann distribution in the gravitational field of Bose star. Velocity distribution is given by Maxwell distribution. We can add molecular matter terms $E_m$ to the energy $E$:
\begin{equation}
\begin{aligned}
E_m=&a_4 N_m \frac{5}{2} kT(r) - a_5 N_M N \frac{Gm_m m}{r}
\\&- a_6 \frac{N_m(N_m-1)}{2} \frac{Gm_m^2}{r}\,.
\end{aligned}
\end{equation}
If the total mass of the molecular cloud bound to the Bose star, $M_m=N_m m_m $, is relatively small as compared with the mass of the Bose star, $N_m m_m \ll Nm $, we may assume that in the leading approximation molecular distribution does not affect Bose star parameters. Then we may include gravitational effect of molecules using iterations. 

Time of formation of Bose stars strongly depends on the DM particle mass $m$. For $m >  10^{-22}$ eV the formation time is smaller than the lifetime of the Universe 13.7 billion years (see e.g.\ \cite{Levkov2018}). 

Total mass of the Bose star $M$ is strongly model dependent and depends on time. There are estimates of a  typical scale for expected mass $M$. The QCD axion dark matter field with axion mass $m \sim 10^{-5}$  eV makes  Bose star with mass  $M \sim  10^{-15} M_\odot$  \cite{Levkov2018,Eggemeier2019,Chen2020,Gorghetto2024}. Fuzzy dark matter with particle mass $ m \sim 10^{-22}$  eV may compose a Bose star with mass $M \sim  10^{8} M_\odot$  \cite{Schive2014,Veltmaat2016}. Bose star accumulating  certain critical mass becomes unstable, collapses  and decays to relativistic jets \cite{VISINELLI201864}.
 
Axions and scalars are neutral, so in zero approximation Bose star is transparent and is hardly observable by telescopes.  In principle, Bose stars may be observed using gravitational lensing of passing light and axion decays to photons. We discuss a different possibility: Bose star may  capture atoms, molecules and dust. Radiation and absorption of light by this matter could make Bose stars visible. 

\section{Variational approach to the Bose star solution}

The variational approach allows one to get an approximate analytic solution to the particle density which minimizes total energy of the system of interacting particles. We consider $N$ bosonic DM particles with position vectors $\vec r_i$ and momenta $\vec p_i$. Given that each particle has mass $m$, they can interact gravitationally. The corresponding non-relativistic Hamiltonian is
\begin{equation}
    H  = -\frac{\hbar^2}{2m} \sum_{i=1}^N \nabla_k^2 - \sum_{i<j}^N \frac{G m^2}{r_{ij}}\,,
    \label{H}
\end{equation}
where $G$ is the gravitational constant and $r_{ij}\equiv |\vec r_i - \vec r_j|$. 

In the ground state, all particles are in the same state with the wave function $\psi(\vec r_i)$, and the full system is described by a product of these wave functions, 
\begin{equation}
\Psi = \prod_{i=1}^N \psi(\vec r_i)\,.
\label{Psi}
\end{equation}
Below we consider two approximate analytic solutions to the function $\psi(\vec r)$.

\subsection{Exponential ansatz}

The ground state wave function may be searched for within the ansatz
\begin{equation}
    \psi(r) = \frac1{\sqrt{\pi}} R^{-3/2}e^{-r/R}\,,
    \label{psiSolution}
\end{equation}
where $R$ is a free parameter. The function $\psi(r)$ is the 1s solution of the stationary Schr\"odinger equation with the potential $V = -\frac{\hbar^2}{m Rr}$ and energy $E_0 = -\frac{\hbar^2}{2 mR^2}$.

Making use of the function (\ref{Psi}) we find the expectation value of the Hamiltonian (\ref{H}):
\begin{equation}
    E = \langle \Psi|H|\Psi \rangle 
    =\frac{N}{2}\left[
    \frac{\hbar^2}{m R^2} - \frac58 \frac{Gm^2}{R} (N-1)
    \right].
\end{equation}
This energy reaches its minimum value 
\begin{equation}
    E_\text{min} = -\frac{25}{512} \frac{G^2 m^5}{\hbar^2} N(N-1)^2
    \label{Emin}
\end{equation}
at
\begin{equation}
    R = \frac{16}{5}\frac{\hbar^2}{Gm^3 (N-1)}\,.
    \label{Zeff}
\end{equation}

The energy of the solution (\ref{Emin}) agrees at 90\% level with the corresponding numerical result in Ref.~\cite{Ruffini1969}.\footnote{Numerical solution of the Schr\"odinger-Poisson equation for the gravitating system on $N$ bosons was first presented in Ref.~\cite{Ruffini1969}. The energy of this system in the large $N$ limit was calculated to $E=- 0.1626 N^3 G^2 m^5\hbar^{-2}$ which was factor 3 off the correct result $E=- 0.05426 N^3 G^2 m^5\hbar^{-2}$ as was pointed out in Ref.~\cite{HartreeSolution}. The 10\% discrepancy of our solution (\ref{Emin}) from the numerical result is still acceptable accuracy for qualitative study of this system considering the simplicity of the trial function (\ref{psiSolution}).  For the Gaussian ansatz (\ref{psi-Gaussian}) the error is 5 times smaller.}

The root-mean-squared (RMS) radius of this solution is
\begin{equation}
    r_\text{rms} = \left(\langle \Psi | r_i^2 | \Psi \rangle \right)^{1/2}
    =\sqrt3 R\,.
\end{equation}
This quantity characterizes the size of the Bose star. We will use $r_\mathrm{rms}$ as the effective boundary of the Bose star in subsequent sections.

As an example, consider a Bose star with mass of Bennu asteroid, $M=7\times 10^{10}$~kg composed of QCD axions with mass $m=10^{-5}$\,eV. It consists of $N=4\times 10^{51}$ particles. Its size is
\begin{equation}
    r_\text{rms} = 4\times 10^{10}\,\mbox{km}.
\end{equation}
Note that the diameter of the Bennu asteroid is 0.5 km.

The particles in the obtained solution have a distribution of velocities. An important characteristic of this distribution is the root-mean-squared speed, which represents a typical speed of particles in the Bose star,
\begin{equation}
    v_\text{rms} = \frac1m \left(
    \langle \Psi |\vec p_i^2| \Psi \rangle 
    \right)^{1/2} =\frac{\hbar}{m R}\,.
\end{equation}

For the axion star with mass $M=7\times 10^{10}$ considered above, the RMS speed is $v_\mathrm{rms}\approx 8\times 10^{-16}c$. Thus, the non-relativistic approximation is justified. However, for solutions with boson  mass $m \sim 10^{-5}$\,eV and total mass of order the Solar mass, a relativistic treatment is needed.


\subsection{Gaussian ansatz}

The simple exponential ansatz (\ref{psiSolution}) for the wave function provides a rather low accuracy for the energy. A better accuracy may be achieved within the Gaussian ansatz \cite{GaussianAnsatz}
\begin{equation}
    \psi(r) = \left( \frac{2}{\pi R^2} \right)^{3/4}
    e^{-r^2/R^2}\,,
    \label{psi-Gaussian}
\end{equation}
where $R$ is the parameter to be found. Making use of this wave function, we find the expectation value of the Hamiltonian (\ref{H}):
\begin{equation}
    E = \langle H \rangle = \frac{3N\hbar^2}{2m R^2} - \frac{N(N-1)Gm^2}{\sqrt{\pi}R}\,.
\end{equation}
This energy reaches its minimum value 
\begin{equation}
    E_\text{min} = -\frac{N(N-1)^2}{6\pi}\frac{G^2 m^5}{\hbar^2}
    \label{EGaussian}
\end{equation}
at
\begin{equation}
    R = \frac{3\sqrt{\pi}\hbar^2}{(N-1)Gm^3}\,.
    \label{Rmin}
\end{equation}

The energy (\ref{EGaussian}) agrees at 98\% level with the numerical solution presented in Ref.~\cite{Ruffini1969} (see the footnote on page 2). Therefore, we will consider only the Gaussian ansatz in what follows.

The RMS radius calculated with the wave function (\ref{psi-Gaussian}) is 
\begin{equation}
    r_\text{rms} = \frac{\sqrt3 R}{2} = \frac{3\sqrt{3\pi}\hbar^2}{2(N-1)Gm^3}\,,
    \label{r-rms-Gaussian}
\end{equation}
and the RMS speed is
\begin{equation}
    v_\text{rms} =\frac{\sqrt3 \hbar}{m R}\,.
    \label{vrms}
\end{equation}

The non-relativistic approximation is applicable for  $v_\mathrm{rms}\lesssim 0.1 c$. Making use of equations (\ref{Rmin}) and (\ref{vrms}), this relation translates into the constraint on the particle number and mass,
\begin{equation}
    Nm^2<\frac{0.1 
    \hbar c\sqrt{3\pi}}{G}\approx 1.5 \times 10^{-10}\,\mathrm{g}^2\,.
\end{equation}


\subsection{Gravitational potential}

The Bose star solution is given by the product of $N$ identical wave functions (\ref{psi-Gaussian}) each describing a particle with mass $m$. They create a gravitational potential $V(r)$ obeying the Poisson equation
\begin{equation}
    \nabla^2 V(r) = -4\pi G \rho(r)\,,
    \label{EquationPotential}
\end{equation}
where the mass density is
\begin{equation}
    \rho = N m \psi^\dag \psi
    =N m\left( \frac{2}{\pi R^2} \right)^{3/2}
    e^{-2r^2/R^2}\,.
\end{equation}
The solution to Eq.~(\ref{EquationPotential}) reads
\begin{align}
    V(r) &= - G N m \frac{\mathrm{erf} (\sqrt2 r/R)}{r} \nonumber\\
    &\approx -GNm\left\{
    \begin{array}{ll}
    \frac{2\sqrt2}{\sqrt{\pi}R}
    -\frac{4\sqrt{2} r^2}{3\sqrt{\pi}R^3}
    \quad & (r\ll R)\,, \\
    \frac1r & (r\gg R)\,.
    \end{array}
    \right.
\end{align}

On the boundary $r=r_\mathrm{rms}$, the gravitational potential is
\begin{equation}
    V(r_\mathrm{rms}) = -\frac{2G Nm}{\sqrt3 R}\mathrm{erf}\left( \sqrt{3/2} \right)
    \approx - 0.2 N^2 \frac{G^2 m^4}{\hbar^2}\,.
\label{V(rms)}
\end{equation}
Thus, the escape velocity on the boundary is
\begin{equation}
    v_\mathrm{esc} = \sqrt{2|V(r_\mathrm{rms})|}
     \approx 0.6 NGm^2/\hbar\,. 
     \label{vesc}
\end{equation}

Consider, for example, the QCD axion with mass $m=10^{-5}$ eV and the total mass comparable with the mass of the Earth, $Nm = M_{\oplus}$. The size of this object is about $r_\mathrm{rms} \approx 0.4$ m. The corresponding escape velocity at the boundary is $v_\mathrm{esc}\approx 0.14 c$.

If the DM particle mass is $m=10^{-9}$ eV, the corresponding RMS radius and escape velocity are $r_\mathrm{rms} \approx 6.3 R_\oplus$, $v_\mathrm{esc}\approx 4.3$\,km/s.

\subsection{Adding $\lambda\phi^4$ selfinteraction}

The scalar field $\phi$ is related to the wave function (\ref{Psi}) as
\begin{equation}
    \phi = \frac1{\sqrt{m}}\Psi\,.
\end{equation}
Making use of the Gaussian ansatz (\ref{psi-Gaussian}), we find
\begin{equation}
    \lambda \int \phi^4 d^3r = \frac{\lambda}{\pi^{3/2} m^2 R^3}\,.
\end{equation}
In the contribution to the energy, this expression should be multiplied by $N(N-1)/2$ which accounts for the pairwise particle interaction. Thus, the total energy of this field configuration reads
\begin{subequations}
\label{EnergyLambda}
\begin{align}
    E &= E_\text{kin} + E_\text{grav} + E_\lambda \,,\\
    E_\text{kin} &= \frac{3N\hbar^2}{2m R^2},
\label{Egrav}\\
    E_\text{grav} &= - \frac{N(N-1)Gm^2}{\sqrt{\pi}R}\,,\\
    E_\lambda &=  \frac{N(N-1)}{2} \frac{\lambda \hbar^3}{\pi^{3/2} c m^2 R^3}\,.
\label{Elambda}
\end{align}
\end{subequations}
This expression reaches its extrema at
\begin{equation}
    R_\pm = \frac{3\sqrt{\pi}\hbar^2}{2G m^3(N-1)}
    \left( 
        1\pm \sqrt{1+\frac{2\lambda G m^2(N-1)^2}{3\pi^2\hbar c}}
    \right).
    \label{Zsol}
\end{equation}

The case $\lambda<0$ corresponds to unstable solution because the energy (\ref{EnergyLambda}) is unbounded from below in this case. However, there is a potential barrier separating energy minimum at $R_+$ in Eq.~(\ref{Zsol}) from the collapse area close to $r=0$, with the barrier maximum at $R_-$. In the case of small negative $\lambda$, $|\lambda|\ll \pi^2 \hbar c/[Gm^2(N-1)^2]$, we obtain minimum of energy
\begin{equation}
    E_\text{min} = -\frac{N(N-1)^2}{6\pi}\frac{G^2 m^5}{\hbar^2} + O(\lambda)
    \label{EminN}
\end{equation}
at
\begin{equation}
\label{R33}
    R = \frac{3\sqrt{\pi}\hbar^2}{G m^3(N-1)}
    +\frac{\lambda\hbar(N-1)}{2\pi^{3/2}mc}+O(\lambda^2)\,,
\end{equation}
and maximum at small $R \propto \lambda$,
\begin{equation}
     R = -\frac{\lambda\hbar(N-1)}{2\pi^{3/2}mc} + O(\lambda^2)\,.
\end{equation}
In the case of small $\lambda$ this barrier will be sufficiently  high to make  the   decay time of the  metastable Bose star  exponentially  long, making this kind of Bose star potentially observable. This result is similar to the gravitational instability of scalar fields discussed in Ref.~\cite{Khlopov85}.

For positive selfinteraction constant, $\lambda>0$, only ``+'' solution in Eq.~(\ref{Zsol}) is physical. If $\lambda$ is sufficiently small, the leading contribution has the same form as in Eq.~(\ref{R33}).
For strong selfinteraction, $\frac{2\lambda G m^2(N-1)^2}{3\pi^2\hbar c}\gg1$,
the parameter $R$ is independent of $N$,
\begin{equation}
\label{Rstrong}
    R = \frac{\hbar^{3/2}}{m^2}\sqrt{\frac{3\lambda}{2\pi cG}}\,.
\end{equation}

For an illustration of the above results, let us introduce an unitless energy of the solution $\hat E = E / E_0$, where $E_0 = m N c^2$ is rest energy of the Bose star. Making use of Eqs.~(\ref{EnergyLambda}) this unitless energy may be represented as follows
\begin{equation}
\label{Ehat}
    \hat E = \frac{3}{2}\frac{1}{\hat R^2} - \frac{\hat G}{\sqrt{\pi}\hat R} + \frac{\hat\lambda}{2\pi^{3/2}\hat R^3}\,,
\end{equation}
where we have introduced the following unitless parameters:
\begin{equation}
    \hat R = \frac{mcR}{\hbar}\,,\quad
    \hat G = \frac{NG m^2}{\hbar c}\,,\quad
    \hat\lambda = N \lambda\,.
\end{equation}

\begin{figure}
    \centering
    \includegraphics[width=\linewidth]{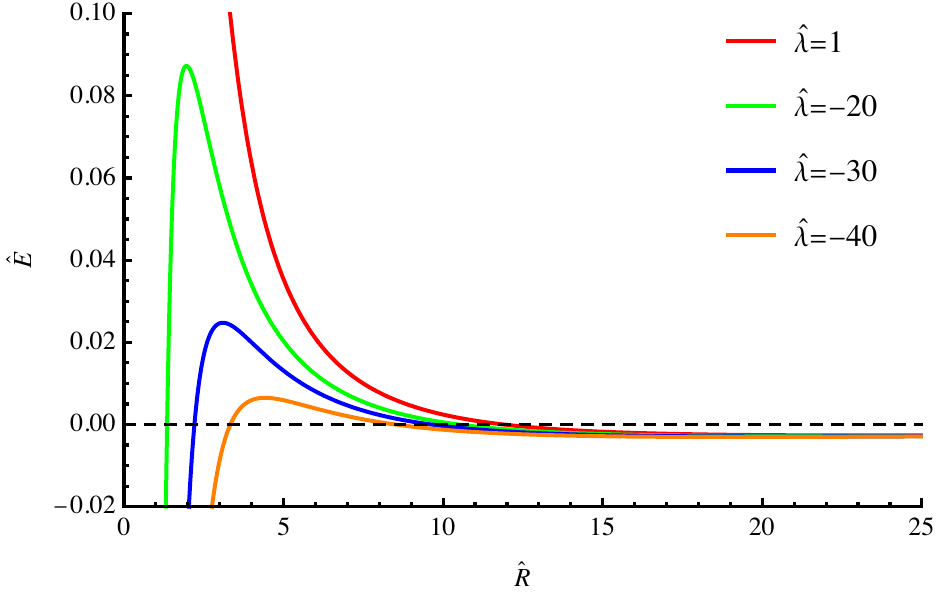}
    \caption{Plot of the unitless energy $\hat E$ as a function of the unitless size parameter $\hat R$ for different values of the selfinteraction constant $\hat\lambda = N\lambda$. This function has a shallow minimum near $\hat R = 24$ and a finite potential barrier for negative $\hat \lambda$.}
    \label{fig:Ehat}
\end{figure}

For instance, consider a Bose star solution with total mass $M=M_\oplus$ composed of particles with mass $m=10^{-5}$\,eV. In this case, $\hat G\approx 0.2$, and the typical behaviour of the unitless energy (\ref{Ehat}) is presented in Fig.~\ref{fig:Ehat} for various values of $\hat\lambda$ ranging from -40 to 1. This function has a shallow minimum near $\hat R_\mathrm{+}\approx 3\sqrt{\pi}/ \hat G\approx 24$, and a finite potential barrier for negative $\hat\lambda$. For positive $\hat\lambda$ this barrier becomes infinitely high.

Note that for the QCD axion, $\phi$, the potential reads $V(\phi)=m^2f_a^2[1-\cos(\phi/f_a)] = \frac12m^2 \phi^2 -\frac{m^2}{4! f_a^2}\phi^4+\ldots$, where $f_a$ is the axion decay constant. The latter is related to the axion mass, $f_a/\mbox{TeV}\simeq 6\mbox{\,keV}/m$, and the values $f_a\lesssim 10^9$\,GeV are excluded, see, e.g., \cite{Sikivie24}. Thus, for the QCD axion, the selfinteraction constant is small and negative, 
\begin{equation}
    \lambda_\mathrm{QCD} = -\frac{m^2}{4! f_a^2}\simeq -10^{-53}\,.
\end{equation}
The full axion potential $V(\phi)$ is, however, non-negative.

\section{Ideal gas distribution around the Bose star}

Let us consider an ideal gas with particles of mass $m_\mathrm{gas}$ in the gravitational potential $V(r)<0$ at temperature $T$. Assuming that the total mass of the gas is sufficiently small so that the gravitational potential is not changed significantly, we can write the density distribution in the gas cloud,
\begin{equation}
    \rho(r) = \rho_0 \exp\left( \frac{m_\mathrm{gas} V(r)}{k_\mathrm{B}T} \right)\,.
\end{equation}
The total mass of the gas cloud is
\begin{equation}
    M_\mathrm{gas} = 4\pi \rho_0 \int_0^\infty 
    \exp\left( \frac{m_\mathrm{gas} V(r)}{k_\mathrm{B}T} \right) r^2 dr\,.
\end{equation}

The gas at every point should follow the Maxwell-Boltzmann distribution
\begin{equation}\label{Maxwell}
    f(v) = 4\pi v^2 \left( \frac{m_\mathrm{gas}}{2\pi k_\mathrm{B}T} \right)^{3/2} e^{-\frac{m_\mathrm{gas} v^2}{2 k_\mathrm{B} T}}\,.
\end{equation}
This distribution, however, should be truncated at $v=v_\mathrm{esc}$. Note that $v_\mathrm{rms} = \sqrt{3k_\mathrm{B} T/m_\mathrm{gas}}$.

At the Earth's surface, the temperature is about $T=300$ K. Empirically, we find the condition for the gas molecules to be captured by the gravitational field of the Earth:
\begin{equation}
\label{v-condition}
    v_\mathrm{rms} \lesssim 0.1v_\mathrm{esc}\,.
\end{equation}
This corresponds to the escape probability suppressed by the  exponential factor in the distribution (\ref{Maxwell}) $\exp{(-{\frac32\frac{v_\mathrm{esc}^2}{v_\mathrm{rms}^2}})} \lesssim \exp{(-150)}$. For the Earth, this condition implies that the gas molecules with mass number $A<6$ escape from the gravitational attraction. This is confirmed by the fact that the fractions of hydrogen and helium gases in the Earth's atmosphere are very low.

Below we extend the condition (\ref{v-condition}) to Bose star solutions. We will consider separately the cases of pure gravitational interaction and a strong selfinteraction between DM particles.

\subsection{Pure gravitational interaction}

In the case of Bose stars, making use of Eq.~\eqref{vesc} the condition (\ref{v-condition}) may be cast in the form
\begin{equation}
    m_\mathrm{gas} \gtrsim\frac{830k_\mathrm{B}T\hbar^2}{N^2 G^2 m^4}\,,
\end{equation}
Representing the gas particle mass as $m_\mathrm{gas} = A m_p$, with $m_p$ the proton mass, we find the condition on the atomic number $A$ of gas particles captured near the surface of the Bose star,
\begin{align}
    A\gtrsim&\frac{830k_\mathrm{B}T\hbar^2}{N^2 G^2 m^4 m_p}\nonumber\\
    &= 1.35 \cdot 10^{-29}\left(\frac{M_\odot}{M}\right)^2\left(\frac{\mathrm{eV}/c^2}{m}\right)^2
    \left(\frac{T}{10 K}\right)\,.
\label{Aresult}
\end{align}

\begin{figure*}
    \centering
    \begin{tabular}{cc}
    \includegraphics[width=8cm]{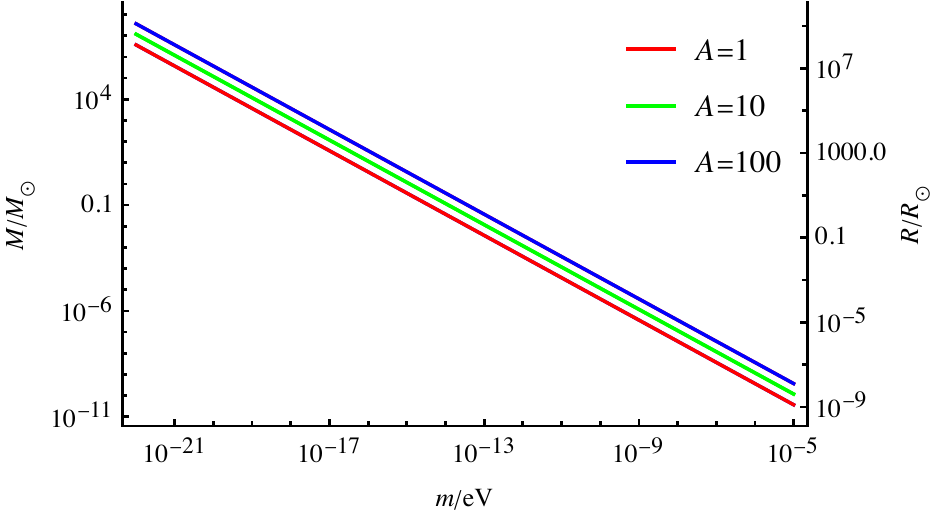}&
    \includegraphics[width=7cm]{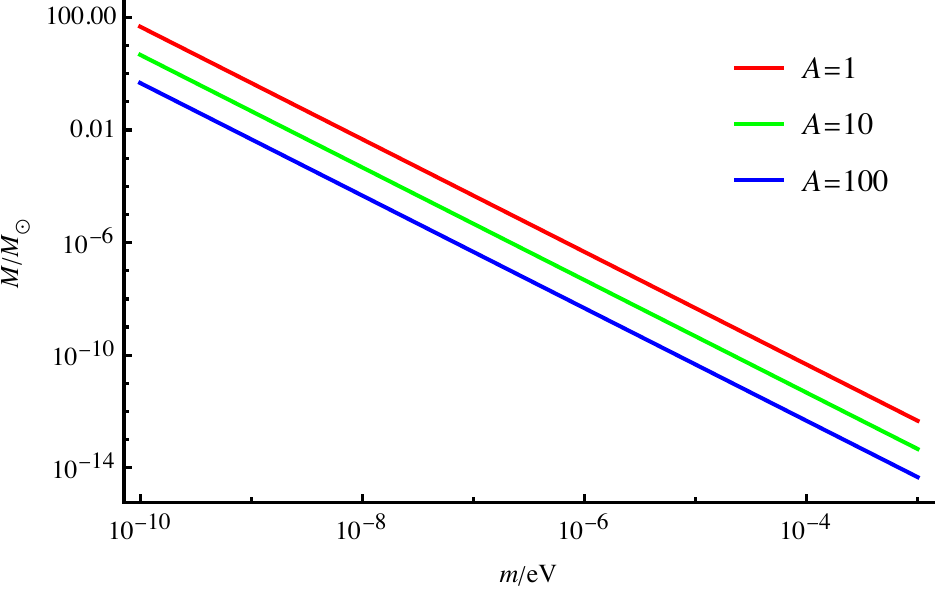}\\
    (a) & (b)
    \end{tabular}
    \caption{Relation between the mass and radius of the Bose star $M$ (in units of solar mass and radius) and the scalar particle mass $m$, assuming that this object can gravitationally bound gas particles with atomic mass numbers $A$ exceeding 1, 10 and 100. The gas is assumed to have temperature $T=10$\,K. Left panel (a): pure gravitational interaction, $\lambda=0$; right panel (b): strong repulsive interaction between the DM particles, $\lambda=10^{-53}$.}
    \label{fig:Aplot}
\end{figure*}

The temperature in cold molecular clouds in our Galaxy is typically in the range from 10 to 20 K. Assuming $T=10$\,K, we estimate the possibility of capture of gases in the fuzzy dark matter with $m=10^{-22}$\,eV and total mass of the object $M=Nm=10^7M_\odot$:
\begin{equation}
    A>13\,.
\end{equation}
Thus, this object can accumulate only heavy atoms and molecules, as well as dust particles. 

In Fig.~\ref{fig:Aplot}a, we present plots of mass (and radius) of the Bose star as a function of scalar particle mass $m$ which can capture molecules with atomic mass numbers $A=1$, 10, and 100.

If radius $R$ of Bose star is small, there  will be too little matter inside the star and impossible to observe it. However, matter may be captured to distant orbits, but this will require bigger mass $M$  of the star, since the escape velocity will be smaller; for matter at distance $r$, the  required star mass is $M(r)=M_0 R/r$, where $M_0$ is the star mass required to capture matter inside. 

Cold molecular gas and dust may also be captured by large axion clusters.

If matter capture conditions  are  fulfilled inside Bose star for all $A$, there will be violation of such conditions outside Bose star at some distance $r$ from the centre. Firstly, this condition will be violated for $A=1$, going further, for higher $A$.  Therefore, there will be separation of heavy and light elements outside the Bose star. This is important for Bose star of large mass and small radius.

The results of this subsection qualitatively apply to the case of attractive self-interaction with small negative $\lambda$. In this case the size parameter $R$ receives a small correction (\ref{R33}) which may be neglected in the result (\ref{Aresult}). The case of strong attractive interaction is not interesting within the present work because it leads to collapse of the boson star to the center and raises the questions of survival of spatially non-compact solutions. 

\subsection{Strong repulsive self-interaction}

In the case of dominant self-interaction with positive coupling constant $\lambda>0$, the size parameter $R$ is given by Eq.~(\ref{Rstrong}). Near the boundary of the Bose star, $r=r_\mathrm{rms}$, the repulsive self-interaction energy (\ref{Elambda}) becomes comparable with the gravitational attraction one (\ref{Egrav}), but it quickly drops with distance by the law $\frac1{r^3}$. Therefore, we can still use Eqs.~(\ref{V(rms)}) and (\ref{vesc}), but with the size parameter $R$ defined by Eq.~(\ref{Rstrong}). As a result, we find the following expression for the escape velocity,
\begin{equation}
    v_\mathrm{esc} \simeq 1.8 N^{1/2} G^{3/4} m^{3/2} \lambda^{-1/4} c^{1/4} \hbar^{-3/4}\,.
\end{equation}
Making use of the empirical relation (\ref{v-condition}) and the ideal gas root-mean-squared velocity $v_\mathrm{rms} = \sqrt{3k_B T/m_\mathrm{gas}}$, we find the condition of the capture of gas particles with mass $m_\mathrm{gas}$:
\begin{equation}
    m_\mathrm{gas} \gtrsim \frac{100 k_B T\sqrt{\lambda \hbar^3}}{G^{3/2}m^3 N \sqrt{c} }\,.
\end{equation}
This relation may be rewritten for the atomic number $A\equiv m_\mathrm{gas}/m_p$,
\begin{equation}
    A \gtrsim 4.7\times 10^{-19}
    \left(\frac{T}{10 K} \right)
    \left(\frac{M_\odot}{M} \right)
    \left(\frac{\mathrm{eV}}{m} \right)^2
    \left(\frac{\lambda}{10^{-53}} \right)^{1/2}\,.
\end{equation}
Note that in contrast with Eq.~(\ref{Aresult}) this relation has different dependence on the total Bose star mass $M$.

In Fig.~\ref{fig:Aplot}b we present plots of mass of the Bose star as a function of scalar particle mass $m$ which can capture molecules with atomic mass numbers $A=1$, 10 and 100 in the case of repulsive scalar particle self-interaction with the coupling $\lambda = 10^{-53}$.


\section{Summary}

We applied the analytic Gaussian solution for Bose star density to study the distribution of interstellar gas particles within the gravitational field of a Bose star, assuming the gas does not significantly perturb the star's gravitational field. We examined the conditions under which interstellar gas particles with a given atomic number $A$ could escape the gravitational attraction of a Bose star with a total mass $M$ composed of scalar dark matter particles of mass $m$. Our findings indicate that, for a range of values of $m$ and $M$, the Bose star can capture atoms, molecules, and dust particles.

A small mass cloud of interstellar gas cannot form a gravitationally bound state without an additional gravitational source. Thus, the observation of such a gravitationally bound gas cloud would suggest it is held by the gravitational field of a Bose star or a dark matter cluster. Due to differences in escape velocities between light and heavy molecules, molecular clouds captured by a Bose star may exhibit higher metallicity compared to typical giant molecular clouds. The discovery of small mass, gravitationally bound molecular clouds with high metallicity in the interstellar medium would provide compelling evidence supporting this dark matter model. To the best of our knowledge, such objects have not been observed so far. The aim of this paper is to motivate the search of compact gas clouds with high metallicity in the interstellar medium of our galaxy. Radio telescopes such as ALMA or GBT and infrared telescopes (e.g., James Webb Space Telescope), which are used to study cold molecular clouds, could be employed in the search for such objects based on new measurements and the analysis of collected data.

\vspace{2mm}
\textit{Acknowledgements.}--- We are grateful to Marco Gorghetto, Maxim Khlopov, Mikhail Kozlov, Sergei Levshakov and John Webb for the interest in this work, useful discussions and references. The work was supported by the Australian Research Council Grants No.\ DP230101058 and No.\ DP200100150.


%
    
\end{document}